\documentclass{elsart3p}
\usepackage{graphicx}
\usepackage{amssymb}
\begin{document}

\begin{frontmatter}

\title{
Pinning, Flux Diodes and Ratchets for Vortices Interacting with Conformal Pinning Arrays
} 
\author[lanl]{C.J. Olson Reichhardt,}
\ead{cjrx@lanl.gov}
\author[nd,anl]{Y. L. Wang,}
\author[anl,niu]{ Z. L. Xiao,}
\author[anl]{ W. K. Kwok,}
\author[lanl]{D. Ray,}
\author[lanl]{C. Reichhardt,}
\author[nd]{B. Jank{\' o}}

\address[lanl]{Theoretical Division,
Los Alamos National Laboratory, Los Alamos, New Mexico 87545, USA}
\address[nd]{Department of Physics, University of Notre Dame, Notre Dame, Indiana
  46556, USA}
\address[anl]{Materials Science Division, Argonne National Laboratory, Argonne, Illinois
 60439, USA}
\address[niu]{Department of Physics, Northern Illinois University, DeKalb, Illinois
  60115, USA} 

\begin{abstract}
  A conformal pinning array can be created by conformally transforming
  a uniform triangular pinning lattice to
 produces a new structure in which the six-fold ordering 
of the original lattice is conserved
but where there is a  spatial gradient in the density of pinning sites.
Here we examine 
several aspects of vortices interacting with conformal pinning arrays and 
how they can be used to create a flux flow diode effect for driving vortices in
different directions across the
arrays. 
Under the application of an ac drive, a  pronounced vortex ratchet effect occurs
where the vortices flow in the easy direction of the array asymmetry.
When the ac drive is applied perpendicular to the asymmetry direction of the
array, it is possible to realize a transverse vortex ratchet effect 
where there is a generation of a dc flow of vortices perpendicular to the ac drive
due to the creation of a noise correlation ratchet by the plastic motion of the vortices.  
We also examine vortex transport  simulations in experiments and compare the pinning
effectiveness
of conformal arrays to uniform triangular pinning
arrays.
We find that a triangular array generally pins the vortices more effectively
at the first matching field and below, 
while the conformal array is more effective at higher fields where
interstitial vortex flow occurs.  
\end{abstract}
\end{frontmatter}

\section{Introduction}
Due to advances in nano-fabrication techniques it is possible 
to create a variety of vortex pinning array geometries in type-II superconductors. 
These include square  \cite{1,2,3,N,4,5},
triangular \cite{6,7,8}, rectangular \cite{9,10,11}, 
honeycomb \cite{12,13}, spin ice \cite{14,15}, 
quasiperiodic \cite{16,17,18,19} and other geometries \cite{20,21}. 
In these systems 
a rich variety of novel vortex crystalline states can form, and
there can be transitions between commensurate and incommensurate states 
as a function of vortex density.
Another motivation to create such arrays is to identify
optimal pinning arrangements that can produce
the highest critical currents or allow for controlled vortex motion for fluxonic devices. 

\begin{figure}
\begin{center}
  \includegraphics[width=\columnwidth]{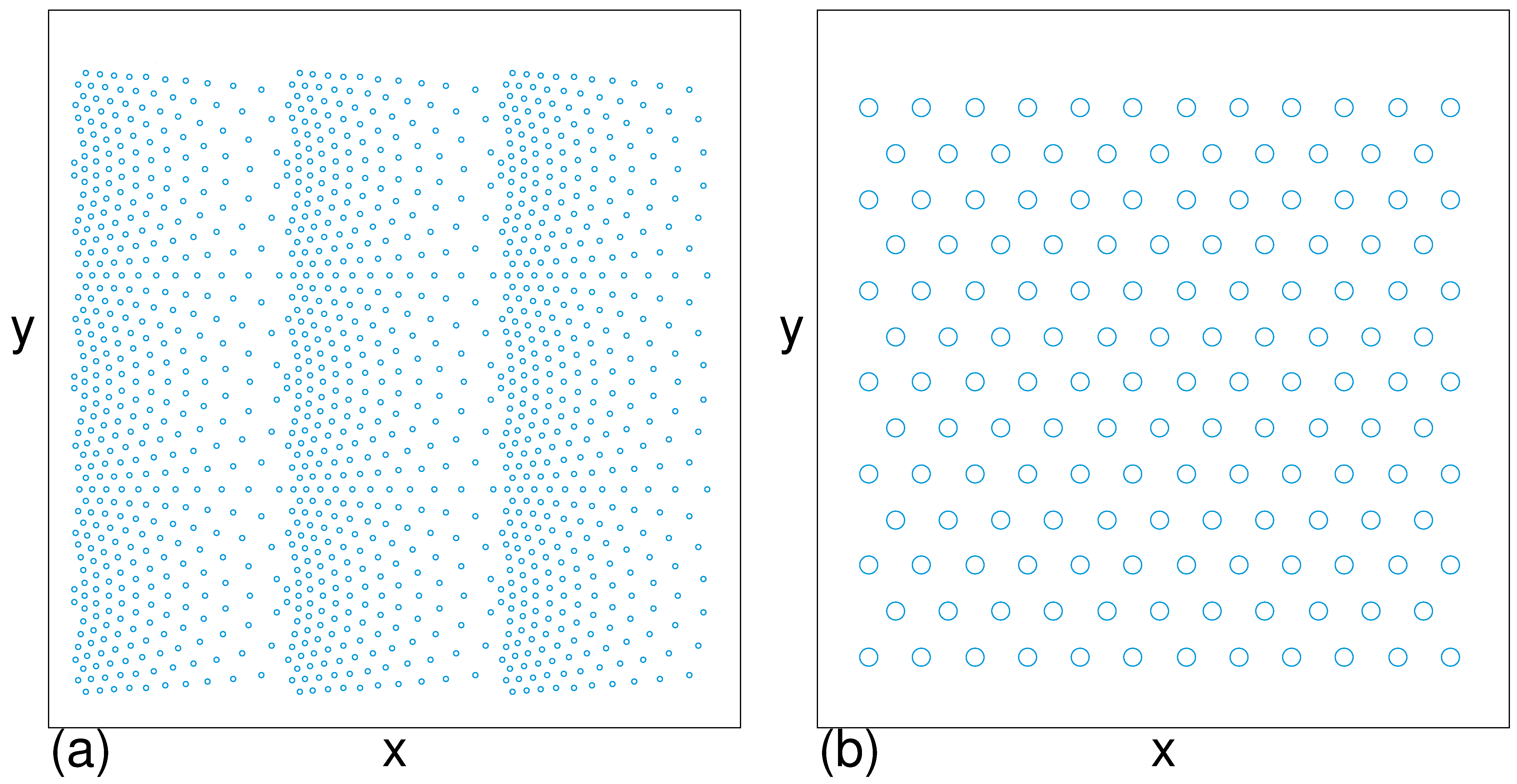}
  \end{center}
\caption{Simulation pinning geometries
  (a) Three conformal crystal pinning arrays placed so as to produce a density
  asymmetry along the $x$-direction. 
(b) An example of a uniform triangular pinning array.
}
\label{fig:1}
\end{figure}

Recently a new type of pinning geometry was 
proposed called a conformal crystal \cite{22,23}.
It is constructed by applying a conformal transformation 
to a uniform triangular array of points 
to obtain a new structure in which the six-fold triangular ordering is preserved but 
where there is a spatial gradient in the density of pinning sites \cite{24}. 
Such arrays potentially have interesting properties for enhancing pinning since 
vortex-vortex interactions naturally favor 
a triangular ordering, while an increasing magnetic field causes the vortices 
to naturally adopt a density or Bean gradient \cite{25,26}. 
There are conditions where the conformal array
is not as optimal as uniform triangular arrays, such as at 
integer matching fields
where the gradient in the conformal arrays can
prevent formation of a completely commensurate state.
Initial simulations 
showed that conformal pinning arrays produced enhanced pinning
compared to uniform random arrays, random arrays with a gradient,
and periodic pinning arrays at non-matching fields,  particularly
at fields where there are more vortices than pinning sites \cite{22,23}. 
Experiments indicated that the       
triangular pinning arrays produced stronger pinning
than the conformal arrays below the first matching
field, 
while the conformal pinning array generated stronger pinning at higher fields \cite{27}.
Other experiments also found enhanced pinning 
by conformal arrays \cite{28}, and there have also been 
other studies of pinning arrays with spatial gradients \cite{29,30}.
In further simulations 
where a series of conformal arrays were placed back to back as illustrated in 
Fig.~\ref{fig:1},
a pronounced vortex ratchet effect occurs when an ac drive is applied 
in the direction of the asymmetry of the array \cite{31}.
The ratchet effect in the conformal arrays is substantially stronger
than that for random gradient arrays and square arrays
with a one-dimensional (1D) spatial gradient that have
the same number of pinning sites \cite{31}.

Here we highlight several
aspects of vortex dynamics and pinning in conformal
arrays.
We show that a flux motion diode effect can occur in which the pinning is
more effective for driving along one direction of the array.
The
conformal arrays also exhibit a strong anisotropy
and have a much higher depinning force for motion along the asymmetry or $x$ direction
than for motion perpendicular to the asymmetry of the array.   
We demonstrate that a longitudinal rocking ratchet effect occurs when
an ac drive is applied along the asymmetry direction,  
and that a novel transverse vortex ratchet effect can occur when the ac drive 
is applied perpendicular to the asymmetry. 
The transverse ratchet effect, which is weaker than the longitudinal ratchet
effect, is an example of a noise correlation ratchet,
rather than a rocking ratchet which occurs for the longitudinal driving.
We show that the effectiveness of the pinning is stronger for uniform
triangular arrays
at low fields when interstitial vortices are absent,
while the pinning is more effective in conformal arrays when interstitial vortices are present. 

\section{Simulations }   
We consider a two-dimensional sample
with periodic boundary conditions in the $x$ and $y$-directions, 
and apply the external magnetic field 
$H$  in the ${\hat z}$) direction.
In our current driven simulations, the
initial vortex positions are obtained by simulated 
annealing.   
The vortex dynamics are
governed by the following overdamped equation of motion:
\begin{equation}  
\eta \frac{d {\bf R}_{i}}{dt} = 
{\bf F}^{vv}_{i} + {\bf F}^{vp}_{i} + {\bf F}^{dc} + {\bf F}^{ac}_{x,y} .
\end{equation} 
Here $\eta$ is the damping constant.
The repulsive vortex-vortex interaction is
${\bf F}^{vv}_{i} = \sum_{j\neq i}F_{0}K_{1}(R_{ij}/\lambda){\hat {\bf R}}_{ij}$,
where $K_{1}$ is a modified Bessel function,  
${\bf R}_{i}$ is the location of vortex $i$,
$\lambda$ is the London penetration depth,
$R_{ij} = |{\bf R}_{i} - {\bf R}_{j}|$,
$ {\hat {\bf R}_{ij}} = ({\bf R}_{i} - {\bf R}_{j})/R_{ij}$, 
$F_{0} = \phi^{2}_{0}/(2\pi\mu_{0}\lambda^3)$, 
$\phi_{0}$ is the flux quantum, and $\mu_{0}$ is the permittivity.    

The system contains $N_{p}$ non-overlapping pinning sites 
each modeled as a parabolic potential  
with a range of $R_{p} = 0.25\lambda$.
The pinning force has the form 
${\bf F}^{ }_{i} =(F_{p}R^{(p)}_{ik}/r_{p})\Theta((r_{p} -R^{(p)}_{ik})/\lambda){\hat {\bf R}^{(p)}}_{ik}$,  
where $\Theta$ is the Heaviside step function, $F_p$ is 
the maximum pinning force,
${\bf R}_k^{(p)}$ is the location of pinning site $k$,
$R_{ik}^{(p)} = |{\bf R}_{i} - {\bf R}_{k}^{(p)}|$, and
$ {\hat {\bf R}_{ik}^{(p)}} = ({\bf R}_{i} - {\bf R}_{k}^{(p)})/R_{ik}^{(p)}$. 
All forces are measured in units of $F_{0}$ and lengths in units of $\lambda$. 

The creation of the conformal pinning geometry is
described in \cite{22,23}. 
Fig.~\ref{fig:1}(a) shows the pinning geometry with 3
back-to-back conformal arrays with
a periodicity of $12\lambda$.
Fig.~\ref{fig:1}(b) shows a representative triangular pinning array; the array
we simulate contains the same number of pinning sites as the conformal geometry
in Fig.~\ref{fig:1}(a).
The term ${\bf F}^{dc}$ represents a dc driving force, 
and we measure the dc velocity per vortex $\langle V\rangle$.  
The ac driving force term is
${\bf F}^{ac}_{x,y} = F^{x,y}_{ac}\sin(\omega T)({\hat {\bf x}},{\hat {\bf y}})$,
where $F^{x,y}_{ac}$ is the ac amplitude. 
We consider both $F^x_{ac}$ where the ac drive is applied in the $x$ direction
parallel to the array asymmetry, as well as $F^y_{ac}$ where the ac drive is
applied in the $y$ direction perpendicular to the array asymmetry, and measure
longitudinal and transverse vortex ratchet effects.
To quantify 
the ratchet effects, we measure the sum of the displacements in the direction 
of the asymmetry of the array,
$X_{\rm net} = N^{-1}_{v}\sum^{N_{v}}_{i=1}(x_{i}(t) - x_{i}(t_{0}))$, 
where $x_{i}(t)$ is the position of vortex $i$ at time $t$ and $x_i(t_{0})$ is the initial
position of the vortex when the ac driving is first applied.
We measure the value of  $X_{\rm net}$ after 25
ac drive cycles to avoid any transient effects.

\section{Diode Effect}

\begin{figure}
\begin{center}
  \includegraphics[width=\columnwidth]{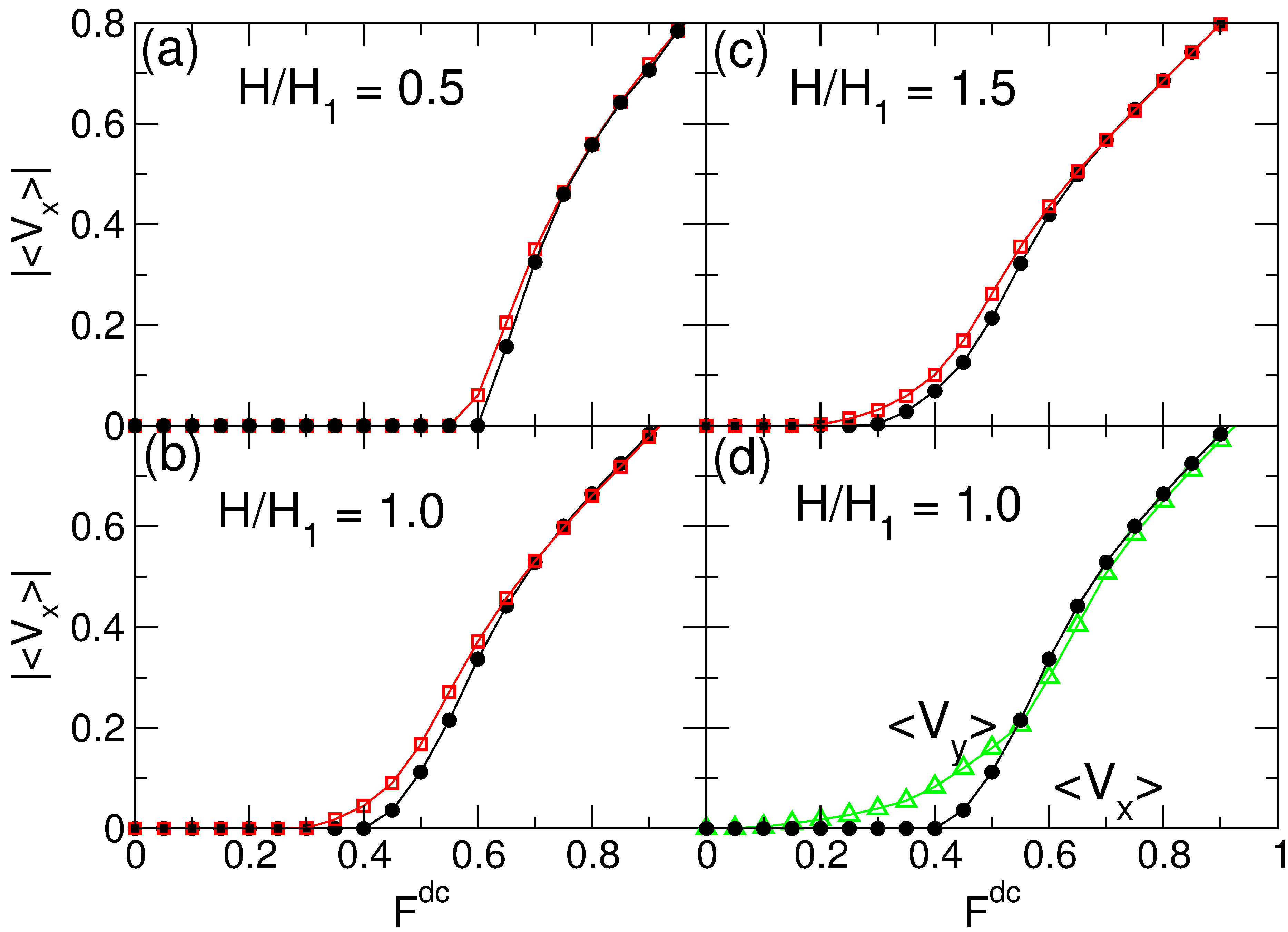}
  \end{center}
\caption{(a-c) The average vortex velocity $|\langle V_{x}\rangle|$
  in the conformal array from Fig.~\ref{fig:1}(a) with $F_{p} = 1.0$
  for driving in the $-x$ direction (open squares) and 
  in the $+x$ direction (filled circles). 
  (a) At $H/H_{1} = 0.5$, the transport curves are asymmetric.
  (b) At $H/H_{1} = 1.0$ the asymmetry is larger.
  (c) At $H/H_{1} = 1.5$ the asymmetry is still present.
  (d) The velocity in the $y$-direction $\langle V_{y}\rangle$  
  for driving in the $+y$ direction of the conformal array (triangles),
  plotted with $\langle V_{x}\rangle$ for driving in the $+x$ direction (filled circules)
  at $H/H_{1} = 1.0$, showing anisotropic transport and a
  crossing of the velocity force curves.
}
\label{fig:2}
\end{figure}

In Fig.~\ref{fig:2}(a,b,c) we plot the absolute value of $|\langle V_{x}\rangle|$  versus
$F^{dc}$
for the conformal array in Fig.~\ref{fig:1}(a)
with $F_{p} = 1.0$ at different fields $H/H_{1}$, where $H_{1}$ is the first
matching field.
The squares denote dc driving in the negative $x$ or easy flow direction, 
and circles denote dc driving in the positive $x$ or hard direction of the array. 
For $H/H_{1} < 0.2$ there is no difference in the depinning threshold for
driving in either direction.
At these low fields, the system is in the single vortex limit,
and since $F_{p}$ is symmetric for an individual pinning site,
the depinning is also symmetric in this limit.  
At $H/H_{1} = 0.5$ in Fig.~\ref{fig:2}(a), there is an asymmetry in the transport  
and the vortices  depin at a lower value of $F^{dc}$ for  driving in the $-x$-direction. 
At $H/H_{1}= 1.0$ in Fig.~\ref{fig:2}(b), this depinning asymmetry is  enhanced,
and within the moving phases the vortex velocities are lower for 
driving in the $+x$ direction.  This aymmetry gradually disappears
for higher $F^{dc} > F_{p}$. 
The  asymmetry persists up to higher values of $H/H_{1}$ as
shown in Fig.~\ref{fig:2}(c) at $H/H_{1} = 1.5$.
The diode effect we observe arises due to collective vortex interactions.  It differs
from other previously reported diode effects for vortices interacting
with asymmetric substrates such as quasi-1D
saw-tooth potentials, where the pinning forces 
have an intrinsic  asymmetry so that a diode effect appears even in the single vortex
limit \cite{32,33}. 

The conformal arrays also exhibit anisotropic transport for driving along
different directions.  In Fig.~\ref{fig:2}(d) we
plot $\langle V_y\rangle$
versus $F^{dc}$ for
a dc drive applied in the $+y$-direction for the system
in Fig.~1(a) at  $H/H_{1} = 1.0$,
along with $\langle V_{x}\rangle$ for driving in the $+x$ direction. 
The $+y$ depinning force
is significantly lower 
than
the $+x$ depinning force;
however, a crossing of the velocity-force curves occurs near
$F^{dc} = 0.5$, above which  the average
vortex velocity for $y$-direction driving becomes lower than
that for $x$-direction driving.
The  lower $+y$ depinning threshold
arises because vortices can more easily depin in the low 
pinning density regions of the sample.
As $F^{dc}$ increases for $+y$ driving, the widths of the regions in which
vortex motion is occurring increases
until $F^{dc} > 1.0$, when all the vortices are moving. 
In contrast, for $+x$ driving,
the vortices cannot move until they are able to pass through the most densely
pinned regions of the sample, so that there is a transition from no vortices moving to
all vortices moving, which produces a higher net average velocity.

\section{Longitudinal and Transverse Ratchets}

The asymmetry of the conformal arrays makes it possible to realize
a ratchet effect
by applying an ac driving force along
the direction of the array asymmetry to produce a
net dc motion of vortices. Previous
vortex ratchet studies focused on cases in which the pinning
substrates or individual pining sites have an intrinsic  asymmetry, such  
as a quasi-1D sawtooth potential \cite{32,33,34}, funnel geometries \cite{35,36}, 
arrays of triangular pinning sites \cite{37,38,39}, 
and other geometries with asymmetrically shaped pinning sites \cite{40,41,42}. 
Vortex ratchets have also been studied in 
systems with random pinning gradients \cite{L} as well as periodic lattices
with 1D gradients \cite{Na}; however the ratchet effects for the
conformal arrays are generally more pronounced than for these systems \cite{31}.   

\begin{figure}
\begin{center}
  \includegraphics[width=\columnwidth]{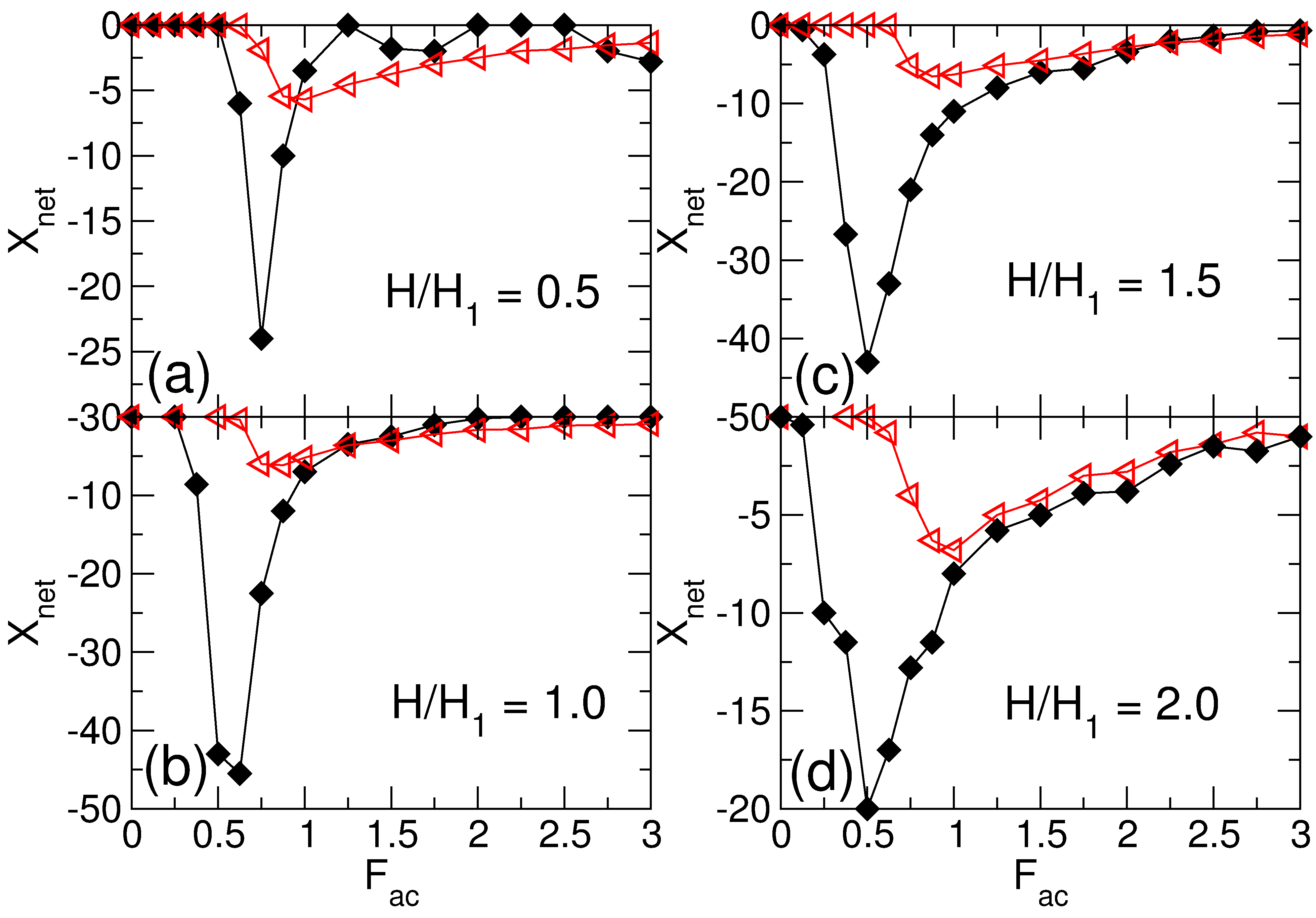}
  \end{center}
\caption{
  $X_{\rm net}$, the particle displacements in the $x$-direction after 25 ac drive
  cycles, for the conformal array at $F_{p}= 1.0$ for ac driving in the $x$-direction
  (filled diamonds) and ac driving in the $y$-direction (open triangles), showing
  both the longitudinal and transverse ratchet effect, at
  $H/H_{1} =$ (a) 0.5, (b) $1.0$ (c), $1.5$ and (d) $2.0$.
}
\label{fig:3}
\end{figure}

In Fig.~\ref{fig:3}(a) we plot the net vortex displacements
$X_{\rm net}$ after 25 ac drive cycles versus ac drive amplitude for
the conformal system with driving in the $x-$ or
$y-$ direction
for $F_{p} = 1.0$ and $H/H_{1} = 0.5$.
The net displacement is in the
$-x$ or easy flow direction
of the substrate.
For the longitudinal drive 
there is a maximum in the ratchet efficiency
near $F^{x}_{ac} = 0.7$ where the vortices move       
an average of $X_{\rm net}=25\lambda$.
For $F^{x}_{ac} > 1.0$ the ratchet effect persists but gradually disappears
at the highest drives. 
When the ac drive is applied in the $y$-direction, Fig.~\ref{fig:3}(a) shows
that there is still
a net drift of vortices in the $-x$-direction, 
indicating the appearance of a transverse ratchet effect in which an
ac drive produces a net dc motion of particles in the direction perpendicular 
to the drive.
Transverse ratchets have been studied for vortices
interacting with triangular shaped pinning sites, 
where the vortices are geometrically deflected by the pinning
site asymmetry
\cite{38,43}.
In the conformal array, the pinning sites are symmetric, so the 
transverse ratchet effect arises from a different process.
Under a $y$-direction ac drive, the vortices undergo plastic motion through
the conformal array,
so that in a co-moving frame the vortices  
experience fluctuations in both the $x$- and $y$-directions. Previous
studies have shown that plastically flowing vortices
exhibit strong non-Gaussian or colored noise fluctuations \cite{44,45}
which, unlike white noise, are time correlated. 
It is known that if a particle is placed in  
an asymmetric substrate and exposed to time-correlated fluctuations,
it will behave as a noise ratchet
\cite{46}.
In the vortex system, the fluctuations
do not arise directly from a fluctuating noise term in  
the equations of motion, but instead originate in
the collective dynamics generated by the plastic flow of the vortices.
Previous studies of interacting particles in a quasi-1D sawtooth substrate 
with added random quenched disorder
showed that the system can exhibit a transverse drift when 
a dc drive is applied in the direction perpendicular to the substrate
asymmetry \cite{47}. The transverse ratchet is generally weaker than
the longitudinal ratchet; however, there are regimes in which
the transverse ratchet effect is comparable
to or even stronger than the longitudinal ratchet effect. 
In Fig.~\ref{fig:3}(b) we show that at
$H/H_{1} = 1.0$, the magnitude of the longitudinal ratchet is larger than at
the lower field of $H/H_{1} = 0.5$.
The transverse ratchet becomes more effective than the 
longitudinal ratchet for $F_{ac} > 1.5$.
At $H/H_{1} = 1.5$ in Fig.~\ref{fig:3}(c) both ratchets effects are present,
and in general,
the maximum efficiency of the longitudinal ratchet decreases
with increasing $H/H_1$ while the transverse ratchet 
efficiency remains almost constant,  as shown in Fig.~\ref{fig:3}(d) at $H/H_{1} = 2.0$.

\section{Pinning Effects}

\begin{figure}
  \begin{center}
  \includegraphics[width=\columnwidth]{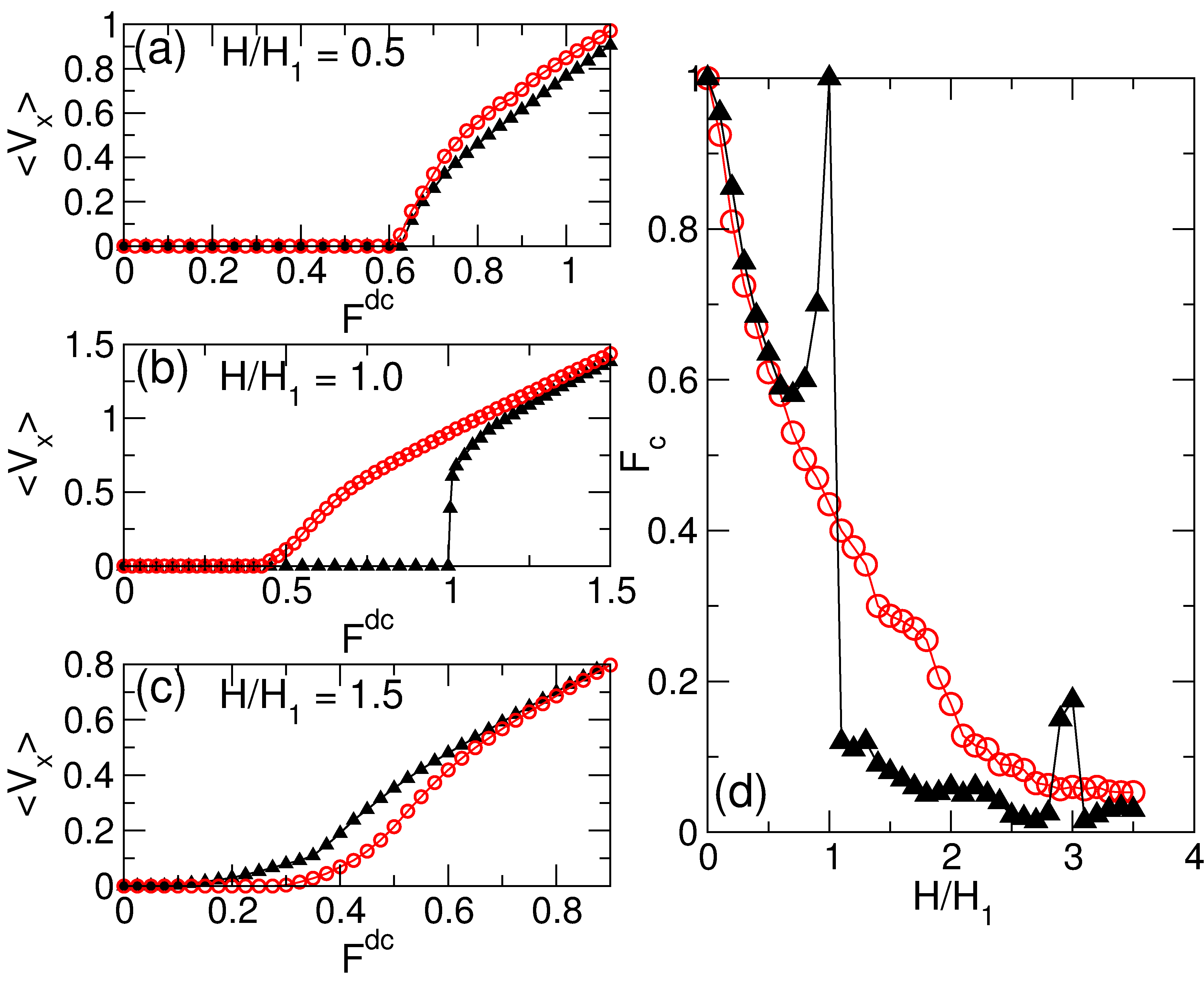}
  \end{center}
  \caption{
    (a,b,c) $\langle V_x\rangle$ vs $F^{dc}$ for driving in the $+x$ direction in the
    conformal array (open circles) and a triangular array (filled triangles) with the same
    number of pinning sites at $H/H_1=$  (a) 0.5, (b) 1.0, and (c) 1.5.
    (d) The depinning force $F_c$ vs $H/H_1$ for driving in the $+x$ direction in the conformal
    array (open circles) and triangular array (filled triangles).
}
\label{fig:4}
\end{figure}

We next compare the dc depinning of the conformal array to that of a uniform triangular
array of the type shown Fig.~\ref{fig:1}(b) with the same number of pinning sites
at $F_{p} = 1.0$.
Previous computational studies focued on the case of flux gradient driven 
dynamics, where the pinning efficiency was determined by
measuring the width of the magnetization hysteresis loops \cite{22,23}. 
Here we study vortex transport with a current driven simulation
in which there is no induced flux gradient, and
we still find strong
differences in the vortex depining for the two cases.
At low fields of $H/H_{1} \leq 0.5$, the system behaves in the single 
vortex limit and the depinning threshold force $F_{c}$
is the same for the triangular and conformal arrays, as shown in
Fig.~\ref{fig:4}(a) where we plot
$\langle V_x\rangle$ versus $F^{dc}$ for driving in the positive $x$
direction for both arrays
at $H/H_{1}=0.5$.
For $0.5 < H/H_{1} \leq 1.0$, $F_{c}$ is higher for the triangular pinning array
as shown in Fig.~\ref{fig:4}(b) at $H/H_{1}=1.0$.
Close to $H/H_{1} = 1.0$,
the vortices in the triangular pinning array form a
commensurate structure in which every pinning site is 
occupied so that the vortex-vortex interactions
effectively cancel and the depinning threshold is
$F_{c} = F_{p}$, while in conformal array only a portion 
of the pinning sites are  occupied due to the
pining gradient so that the depinning force is
less than half of that of the triangular pinning array. 
For $H/H_{1} > 1.0$, interstitial vortices appear in
the triangular pinning array and cause a sharp drop in the depinning threshold,
while $F_c$ for the conformal array remains nearly unchanged and is
now higher than that of the triangular array,
as illustrated in Fig.~\ref{fig:4}(c) at $H/H_{1} = 1.5$. 
In  Fig.~\ref{fig:4}(d) we plot $F_{c}$ versus $H/H_{1}$ for 
the conformal and triangular arrays.   $F_{c}$ is higher for the triangular
array around the matching peaks of  $H/H_{1} = 1.0$ and
$H/H_{1}=3.0$.
For $H/H_{1} > 1.0$ there
is a rapid drop in $F_c$ for the triangular array, and $F_c$ does not increase
again until an ordered commensurate state forms at
$H/H_{1} = 3.0$ \cite{3}. 
We note that the triangular array has only a weak peak in $F_{c}$ 
at $H/H_{1} = 2.0$, where the vortices form a honeycomb state rather than a 
more stable triangular lattice \cite{3}.
For higher values of $F_{p}$, the differences between the two systems is reduced as the
vortex-vortex interactions become less important and the system
transitions to the single vortex pinning regime.
In addition, the drop in $F_{c}$ for the triangular
arrays shifts to higher values of $H/H_{1}$
when the pinning sites are strong enough or
large enough to allow for double or multiple vortex occupancy.  

\section{Sample Fabrication} 

\begin{figure}
  \begin{center}
  \includegraphics[width=\columnwidth]{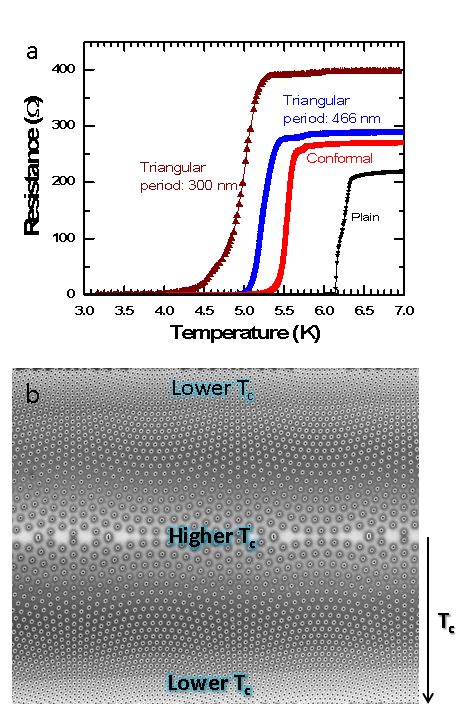}
  \end{center}
  \caption{
    Patterning-induced degradation of superconducting critical temperature.
    (a) R-T curves of focused-ion-beam (FIB) patterned MoGe films with various hole patterns
    (two triangular arrays and a conformal array) and that of a reference plain film. The hole
    sizes in the various patterns are all around 100 nm. The spacing of the two triangular
    arrays are 300 nm and 466 nm, respectively. The $T_c$ is suppressed in the patterned
    sections as compared to the unpatterned plain film and further decreases with shorter
    hole-hole spacing. The conformal hole array has equivalent average hole density with the
    triangular hole array with 466 nm hole-hole spacing and it shows higher $T_c$ than the
    triangular array due to the low density of holes in the center part of the film.
    (b) Non-uniformly distributed $T_c$ in conformal hole array.
}
\label{fig:5}
\end{figure}

In fabricating samples for measurement, we observe that $T_c$ of the sample is
usually suppressed due to processing-induced sample degradation.
Figure \ref{fig:5}(a) shows the R-T curves of a plain MoGe film, two focus-iron-beam (FIB)
patterned MoGe films containing a triangular array of holes with different hole-hole
spacings, and one FIB patterned MoGe film with a conformal array of holes having
an equivalent average hole density as one of the triangular arrays.
All three patterned films have lower $T_c$ values than the unpatterned film,
indicating processing-induced degradation of superconductivity.
Furthermore, the film with a hole-hole spacing of 300nm has
a lower $T_c$ than that with a hole-hole spacing of 460 nm.
That is, when holes are patterned with a process that can obviously degrade the sample,
the sample with a higher density of holes will have a lower $T_c$.
For a non-uniform pinning geometry, such as the conformal pinning array,
the $T_c$ of the sample is spatially inhomogeneous because the hole-hole spacings
vary across the sample.
That is, the region with a higher local density of pinning sites has smaller $T_c$,
as shown in Fig.~\ref{fig:4}(b).  The $T_c$ determined from the transport measurements
gives
values from the current channels that have the highest $T_c$,
resulting in an over-estimate of the true $T_c$. 

Samples with different pinning geometries usually have different $T_c$ values due to
the patterning-induced sample degradation, which confounds the comparison of different
pinning landscapes. One commonly used method for comparing the results from samples
with different $T_c$ is to compare data obtained at the same reduced temperature
$T/T_c$.   Due to the non-uniformly distributed $T_c$ in non-uniform pinning
geometries, however, the reduced temperature $T/T_c$ determined by the transport
method is nearly meaningless. Thus, when studying non-uniformly distributed
pinning geometries, it is important to adopt a sample fabrication method that limits the
damage only to the holes and that does not alter the $T_c$ of the rest of the sample. 

These $T_c$ suppressing effects are frequently observed in FIB patterned samples and
samples fabricated using direct liftoff method. Our sample patterning method of using
E-beam lithography followed by reactive ion etching does not suppress the
$T_c$ of the films \cite{27}.
The film is protected by an E-beam resist mask during the etching process and there is
no damage to the film except for the area of the hole, which not only enables us to
measure and compare the results at the same temperature but also produce holes with
a strong pinning potential.

\section{Experimental Results} 

\begin{figure*}
  \begin{center}
  \includegraphics[width=1.7\columnwidth]{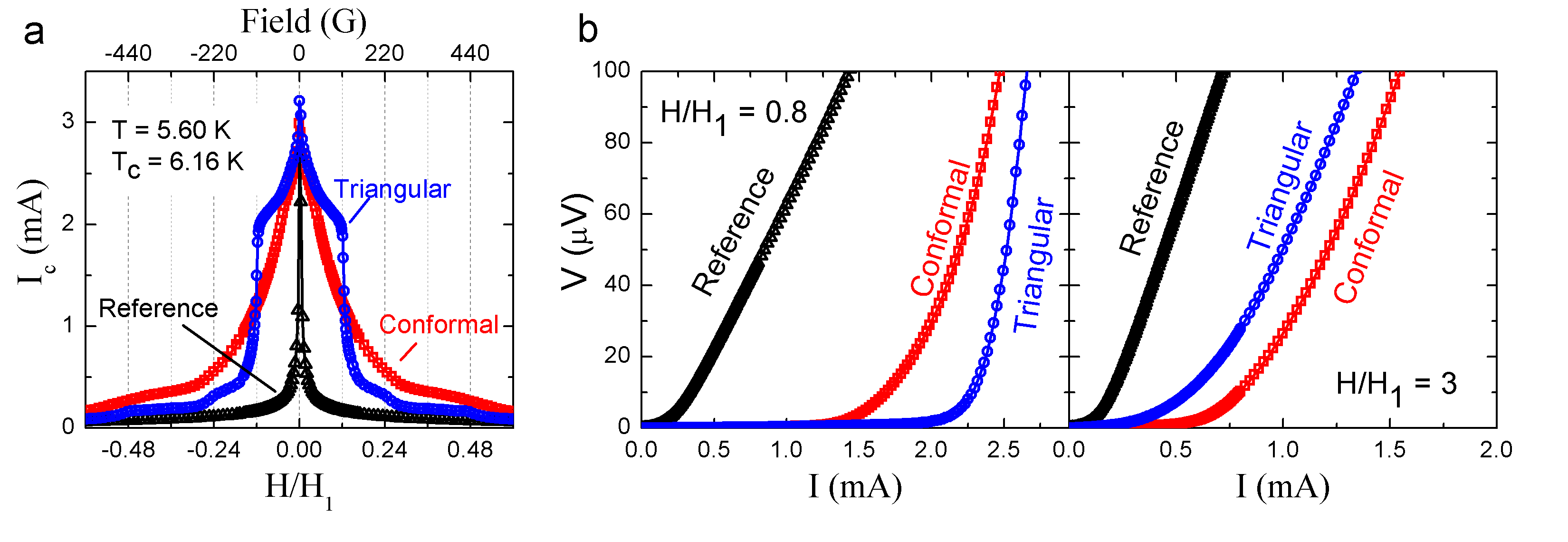}
  \end{center}
  \caption{Pinning performances of a conformal pinscape and a triangular pinscape.
    (a) Critical currents of three sections on the same MoGe micro-bridge with one section
    patterned with a triangular array of holes (hole-hole spacing: 446 nm), one section
    patterned with a conformal array of holes (containing an equivalent average density of
    holes as the triangular array) and one un-patterned reference section.
    (b) I-V curves at magnetic fields of 0.8$H_1$ (left panel) and 3$H_1$ (right panel).
    The data of the three sections are measured at the same time and temperature.
    The $T_c$ of the sample is 6.16 K and the measured temperature is 5.6 K (91\% $T_c$).
}
\label{fig:6}
\end{figure*}

Figure 6(a) shows the magnetic field dependence of the critical current $I_c(H)$
for three sections on the same micro-bridge of a superconducting MoGe film measured
at the same time and temperature (T = 5.6 K). The hole size is about 110 nm which
ensures that each hole can only trap one vortex while additional vortices occupy the
interstitial spaces between holes. Both of the two patterned sections have significantly
enhanced critical currents compared to the unpatterned reference section at all applied
magnetic fields, indicating the pinning effectiveness of the fabricated hole-arrays.
For the triangular pinning array, the steep drop of the critical current curve once the
magnetic field exceeds the first matching value is due to the appearance of interstitial
vortices, which can be depinned at a much lower driving force in a regular pinning array
with easy vortex flow channels.
Above the first matching field, the depinning of the weakly pinned interstitial vortices
makes the third and fourth matchings almost indistinguishable. On the other hand, a
small peak appears at the fifth matching field due to a caging effect of the interstitial
vortices. The $I_c$(H) curve (red) in Fig.~\ref{fig:6}(a) for the section with a conformal
hole array shows a smooth decay in the critical current with increasing magnetic field.
The gradually changing hole density produces localized commensuration at a 
series of local matching fields, 
which smears out the bulk matching field features.
Although these multiple matching length scales can result in pinning enhancement over a
wide range of magnetic field, the incommensuration between the conformal array
and an Abrikosov triangular lattice leads to an increase in the vortex-vortex interaction
energy in the vortex matter, thus reducing the critical current
compared to the triangular hole array, which is perfectly commensurate with the
Abrikosov triangular lattice at magnetic fields lower than the first matching field.
Once interstitial vortices appear above the first matching field, however,
the conformal hole array outperforms its triangular array counterpart in pinning
efficiency.  The pinning performance of different pinning arrays can also be observed
in the I-V curves shown in Fig.~\ref{fig:6}(b).  At the low field $H/H_1=0.8$,
the zero resistance state persists to a higher current for the triangular pinning array,
while at the high field $H/H_3=3$, the conformal pinning array maintains zero resistance
over a larger current range.

\begin{figure}
  \begin{center}
  \includegraphics[width=\columnwidth]{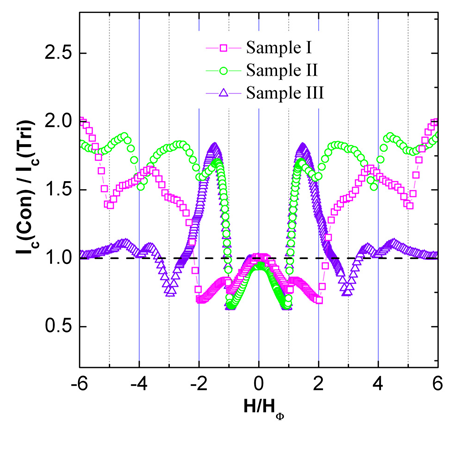}
  \end{center}
  \caption{
    Comparison of critical currents. Ratios of the critical currents for the sections
    with conformal ($I_c$(Con)) and triangular ($I_c$(Tri)) hole arrays of three
    samples with different hole spacing b (average), hole diameters/sizes d and
    hole types (through or blind). Sample I: b=777 nm, d=220 nm, through holes;
    Sample II: b=466 nm, d=110 nm, through holes; Sample III: b=466 nm; d=110 nm,
    blind holes.  
}
\label{fig:7}
\end{figure}

We compare the pinning effect of the conformal and triangular hole arrays of three
different samples in Fig.~\ref{fig:7} by plotting the magnetic field dependence of the
ratios of their critical currents. Sample I and Sample II contain through holes with strong
pinning strength for each individual hole. They have different hole sizes.
Sample I has bigger holes (220 nm in diameter), each of which can trap up to  two
vortices at the measured temperature.  In this sample the performance of the conformal
pinning array surpasses that of the triangular pinning array at fields above the
second matching field. Sample II has smaller holes (110 nm in diameter), each of
which can trap at most one vortex, so the performance of the conformal pinning array
surpasses that of the triangular pinning array above the first matching field.
Sample III also has small holes that are 110 nm in diameter, but instead of through holes,
sample III has blind holes with a much weaker pinning strength for each individual hole.
The performance of the conformal pinning array also surpasses that of the triangular pinning
array above the first matching field. Due to the weaker pinning strength of the holes,
the advantage of the conformal pinning array in sample III at high fields is not as
obvious as in sample II.
In all three samples, the critical currents of the sections containing conformal
hole arrays are lower than those of the sections with holes arranged in a triangular
lattice at low magnetic fields ($H < 2H_1$ for Sample I, $H < H_1$ for Samples II and III)
where all vortices are pinned in the triangular lattice of holes.
At high magnetic fields the conformal array enhances the critical currents in
all three samples except for Sample III which has blind holes and weak pinning strength.
In this case, in a specific field range (around the third matching field in Sample III),
the caging effect of the triangular lattice prevails.

\section{Conclusion}

We have investigated various aspects of vortex motion and pinning in conformal pinning
arrays using both simulations 
and experiments. For dc driving in the direction the asymmetry of the pinning array we
find that there is a diode
effect where the effective pinning is higher in the hard direction of the asymmetry of the
array. 
This pinning asymmetry is a collective effect since in the single vortex limit the diode
effect is absent as the  pinning sites themselves are symmetric.
In the presence of an ac drive in the same direction as the asymmetry of the array, a
pronounced longitudinal vortex ratchet effect can 
occur.  We also show that when the ac drive is applied perpendicular to the array asymmetry
direction, 
a transverse vortex ratchet occurs in which there is a net dc flow of vortices
perpendicular to the ac drive. The longitudinal
ratchet is a realization of of a collective rocking ratchet while the transverse ratchet is
an example of a noise correlation ratchet.
The confomral pinning arrays in general show enhanced pinning over uniform triangular
pinning arrays when the external field is large enough that
interstitial vortices are present, while at fields where commensurate vortex lattice
structures occur, the pining is higher in the triangular pinning arrays.

This work was carried out under the auspices of the 
NNSA of the 
U.S. DoE
at 
LANL
under Contract No.
DE-AC52-06NA25396.

\end{document}